\documentclass[%
pra%
 ,twocolumn%
 ,secnumarabic%
,amssymb, amsmath,nobibnotes, aps, pra]{revtex4}
\usepackage{docs}%
\usepackage{graphicx}
\usepackage{bm}%
\expandafter\ifx\csname package@font\endcsname\relax\else
 \expandafter\expandafter
 \expandafter\usepackage
 \expandafter\expandafter
 \expandafter{\csname package@font\endcsname}%
\fi

\begin{document}

\title{Influence of non-uniformity in sapphire substrates for a gravitational wave telescope}
\author{K. Somiya}
\email{somiya@phys.titech.ac.jp}
\affiliation{Department of Physics, Tokyo Institute of Technology, 2-12-1 Oh-okayama Meguro Tokyo 152-8551, Japan}
\author{E. Hirose}
\affiliation{Institute for Cosmic Ray Research, the University of Tokyo, 5-1-5 Kashiwa-no-ha, Kashiwa, Chiba 277-8582, Japan}
\author{Y. Michimura}
\affiliation{Department of Physics, University of Tokyo, Bunkyo, Tokyo 113-0033, Japan}
\date{July 2019}

\begin{abstract}
Construction of a large-scale cryogenic gravitational-wave telescope KAGRA has been completed and the four sapphire test masses have been installed in cryostat vacuum chambers. It recently turned out that a sapphire substrate used for one of the input test masses shows a characteristic strcuture in its transmission map due to non-uniformity of the crystal. We performed an interferometer simulation to see the influence of the non-uniformity using measured transmission/reflection maps.
\end{abstract}

\pacs{0.00}

\maketitle

\section{Introduction}

Gravitational waves are ripples of the spacetime that transfer information of astronomical events from distant galaxies or from the early universe. Since April 2019, the large-scale telescopes in the US and Europe (Advanced LIGO~\cite{LIGO} and Advanced Virgo~\cite{Virgo}, respectively) have started the third round of the observation run, and the large-scale cryogenic telescope in Japan, KAGRA~{\cite{KAGRA}\cite{KAGRA2}\cite{KAGRA3}}, is to join the observation run from the end of 2019. While the LIGO and Virgo detectors are operated in the room temperature with fused silica test masses, KAGRA will be operated in $20\sim23$\,K with sapphire crystal test masses. The material was chosen mainly for three reasons: one is for its high thermal conductivity at a cryogenic temperature, another is for its high mechanical quality factor at a cryogenic temperature, and the other is the high transmittance for the 1064\,nm laser, which has been used as a light source in a gravitational-wave detector for a few decades. The optical configuration of the telescope is the Michelson interferometer with a Fabry-Perot cavity in each arm. The four test masses in the arm cavities are sapphire mirrors. The beam splitter and other optics are made of fused silica. 
We chose the C-axis of the crystal perpendicular to the mirror surface so that variation of refractive index can be minimum.  
We developed a number of sapphire substrates and selected the two with the lowest optical absorption to be used as the input mirrors of the arm cavities. Note that the requirement on the input mirrors is severer than the end mirrors as the heat absorbed in the input mirror substrate can be as high as the heat absorbed in the coatings. As inhomogeneity of sapphire is not as good as fused silica, 
it is essential to correct the transmitted wavefront curvature by a technique called Ion Beam Figuring (IBF), where physical thickness of the substrate is adjusted point-to-point so that the wavefront error can be within a few nanometers in the clear aperture.
After the IBF treatment, however, we found that the measured figure error is larger than the specification ($<6\,$nm) by a factor of $4\sim5$, and further more there is a ridge-like structure of the height of $\sim100$\,nm in the transmission map, or a so-called transmitted wavefront error (TWE) map, of one of the input mirrors. Our mirror team has concluded that this would have happened because of the use of a circular polarization beam at measurements for the IBF treatment, which is in fact not reasonable if the substrate be the ideal C-axis crystal. The ridge-like structure would be troublesome as the transmitted beam could contain asymmetric modes as if the mirror were tilted. It is necessary to find out the influence of this non-uniformity in the TWE map to some control signals and to the sensitivity of the telescope.

The structure of the paper is as follows. In Sec.~\ref{sec:2} we briefly introduce the interferometer configuration of the KAGRA telescope. In Sec.~\ref{sec:3} we explain inhomogeneity of a sapphire substrate and the non-uniformity in our sapphire test masses more in detail. In Sec.~\ref{sec:4} we introduce a modal-model interferometer simulation code used for the analysis. In Sec.~\ref{sec:5} we show some results of the analysis and discuss the influence.

\section{Interferometer configuration of KAGRA}\label{sec:2}

\begin{figure}[htbp]
	\begin{center}
		\includegraphics[width=8.5cm]{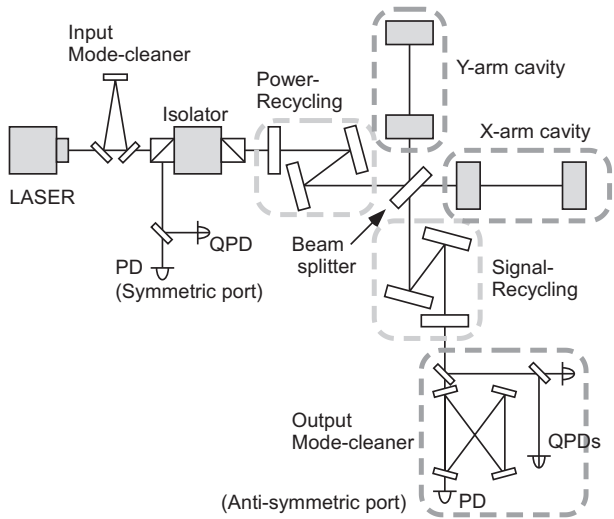}
	\caption{Rough schematics of the interferometer. Some items that are not mentioned in this paper are omitted and the photo-detectors (PDs) and quadrant photo-detectors (QPDs) used in the simulation in this paper are depicted.}
	\label{fig:IFO}
	\end{center}
\end{figure}

Figure~\ref{fig:IFO} shows a schematic of the interferometer. The laser beam in the wavelength of 1064\,nm is first injected to the input mode-cleaner to filter out higher-order spatial modes. The beam is then injected to the input optical isolator, which redirects the beam coming back from the interferometer to protect the laser system. The main interferometer is based on a Michelson interferometer with a Fabry-Perot cavity in each arm and two recycling cavities to effectively enhance the circulation power and the bandwidth. The length and the finesse of the arm cavities are 3\,km and 1550, respectively. Both the input and end mirrors have the radius of curvature of 1.9\,km. The mirror diameter and the thickness are 22\,cm and 15\,cm, respectively, and the mass is 23\,kg. The wedge angle of the input mirror is 0.025\,deg and that of the end mirror is 0.05\,deg. The interferometer is operated in a so-called dark-fringe and all the differential fields including a gravitational-wave signal field go to the anti-symmetric port through the output mode-cleaner. The arm cavities are in fact differentially detuned from the resonance so that a very small fraction of the beam comes out to the anti-symmetric port to be a reference light of the gravitational wave signal. The output mode-cleaner filters out spatial higher order modes and radio-frequency control sidebands while transmitting the gravitational wave signal field and the reference beam. All the output mode-cleaner components are mounted on a suspended breadboard and the relative position and the alignment are controlled using signals from the on-board QPDs. Some more details of the interferometer configuration and the control scheme can be found in~{\cite{KAGRA}\cite{KAGRA4}}.

\section{Inhomogeneity in the sapphire substrate}\label{sec:3}

Prior to the procurement of the actual 23\,kg sapphire test masses, we asked a vendor to polish a test sample, or the {\it pathfinder}, and measured the figure error and the inhomogeneity~\cite{Hirose2014}. The diameter and the thickness of the pathfinder are 200\,mm and 60\,mm, respectively. The measured root-mean-square (RMS) figure error in the central area (140\,mm diameter) was 0.24\,nm, which satisfies our requirement of 0.5\,nm. The measured inhomogeneity of the pathfinder substrate in terms of the error of the refractive index was $\Delta n=1.92\times10^{-7}$, which corresponds to the TWE of 28.8\,nm in a 23\,kg test mass.

The results of the pathfinder tells us a necessity of the IBF treatment to correct the error in the TWE map. We asked the same vendor to polish the 23\,kg sapphire substrates with the IBF treatment for the two substrates that are to be used for the input mirrors (ITMX and ITMY). The specification of the RMS TWE in the central area (140\,mm diameter) was $<6$\,nm and the reported value from the vendor was 3.47\,nm for the ITMX and 4.07\,nm for the ITMY. However, our independent measurement showed a different results: 25.9\,nm for the ITMX and 30.1\,nm for the ITMY. One difference between the measurements at the vendor and our own is that the vendor used a circular polarization beam and our setup used the linear polarization beam in the Fizeau interferometers. 
We concluded that the substrate was not perfectly in the C-axis and sensitive to the polarization to cause an error in the vendor's measurement, which resulted in the high TWE after the IBF treatment.

\begin{figure}[htbp]
	\begin{center}
		\includegraphics[width=4.25cm]{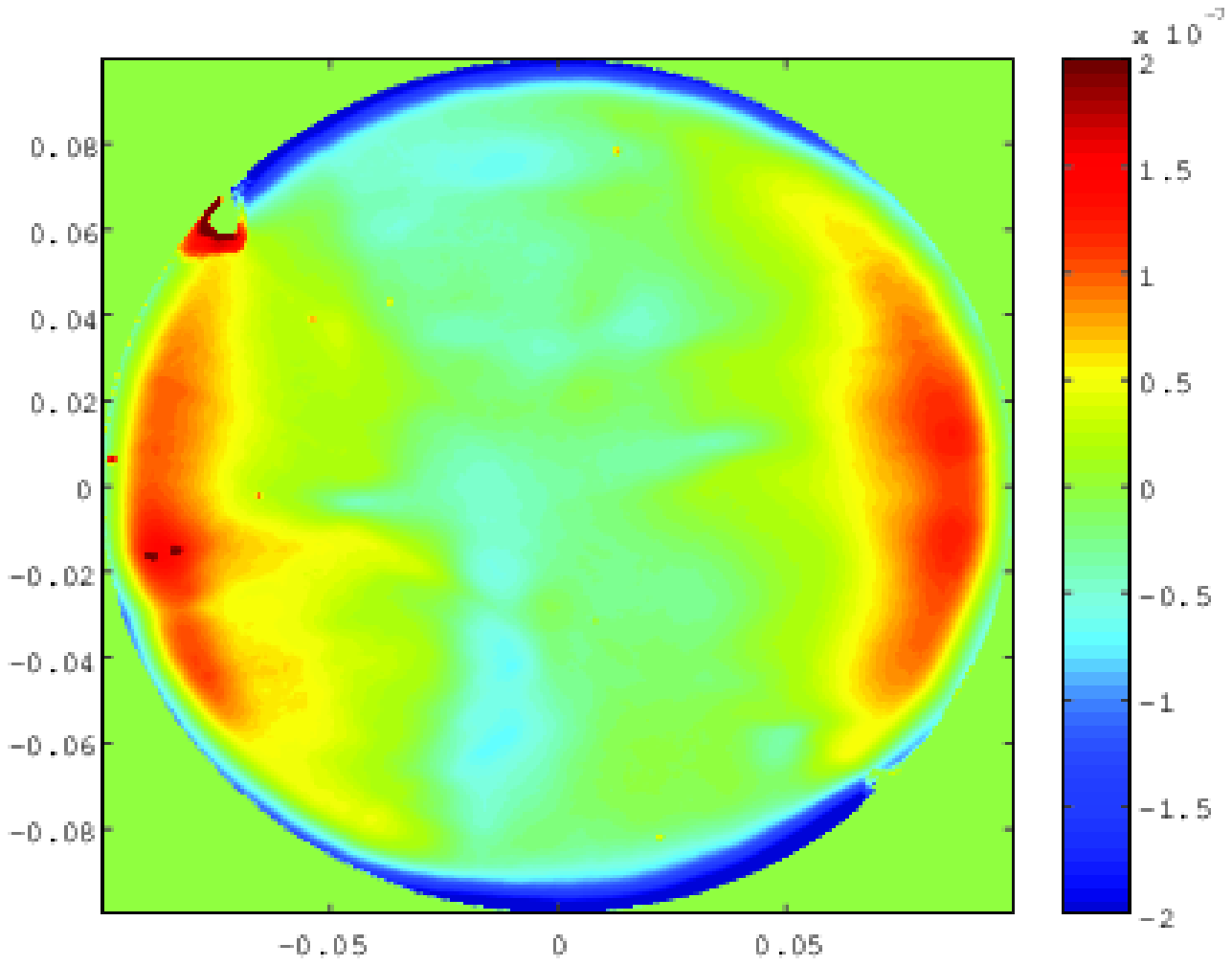}
		\includegraphics[width=4.25cm]{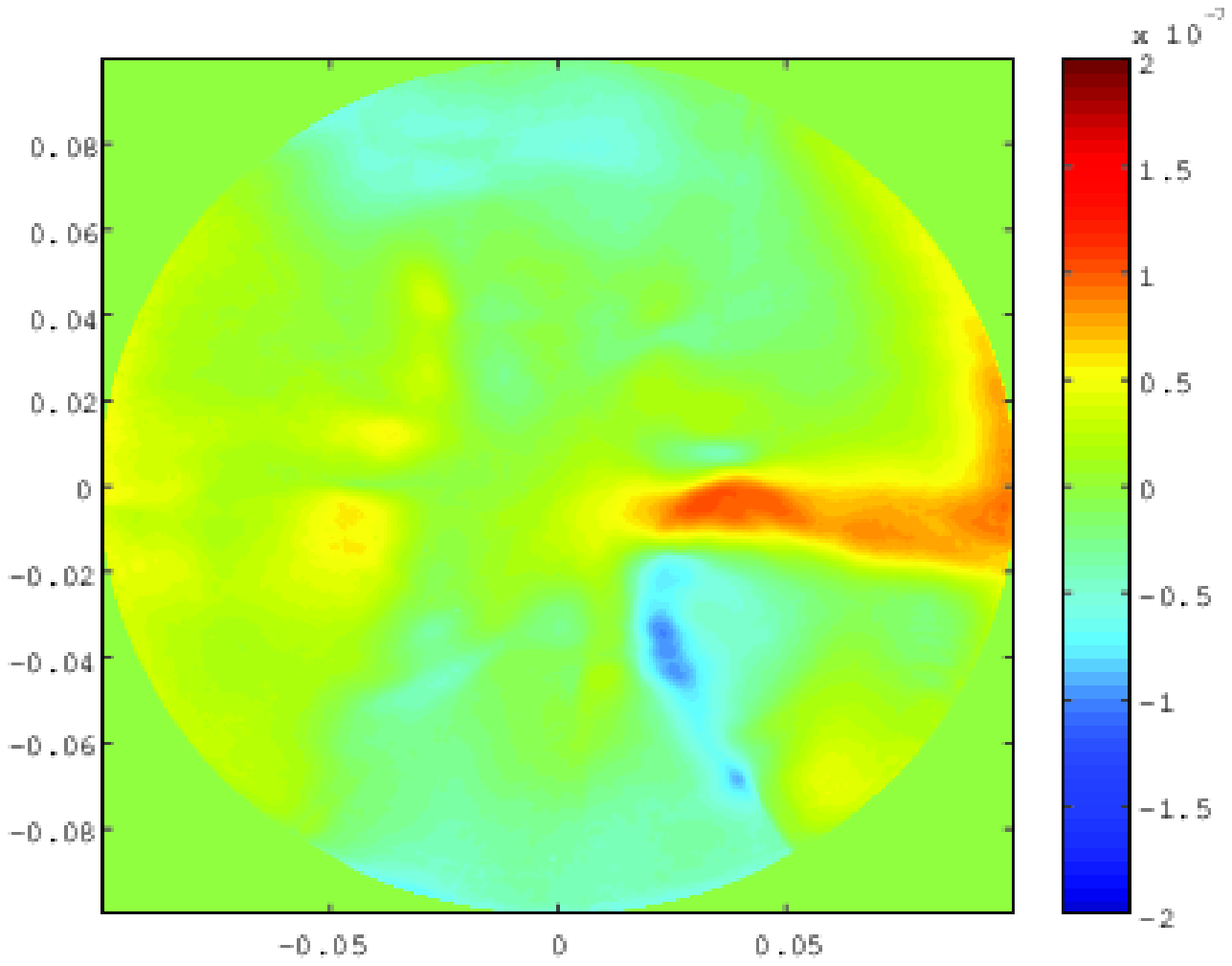}
	\caption{TWE maps of ITMX (left) and ITMY (right) after removing the power term, seen from the other side of the high-reflective surface. The plotted area is 20\,cm in diameter around the center and the color bar ranges in $\pm20\,\mu\mathrm{m}$. The RMS errors in the central area are 25.9\,nm for the ITMX and 30.1\,nm for the ITMY.}
	\label{fig:TWE}
	\end{center}
\end{figure}

Figure~\ref{fig:TWE} shows the measured TWE maps of the input mirrors after removing the power term. The beam radius on the input mirrors is 3.5\,cm so the large errors near the edge in the ITMX would not cause too much trouble while the ridge structure in the ITMY exists near the center and may create a non-trivial coupling between the length motion and yaw motion.

\section{Interferometer simulation with the mirror maps}\label{sec:4}

\begin{figure}[htbp]
	\begin{center}
		\includegraphics[width=4.25cm]{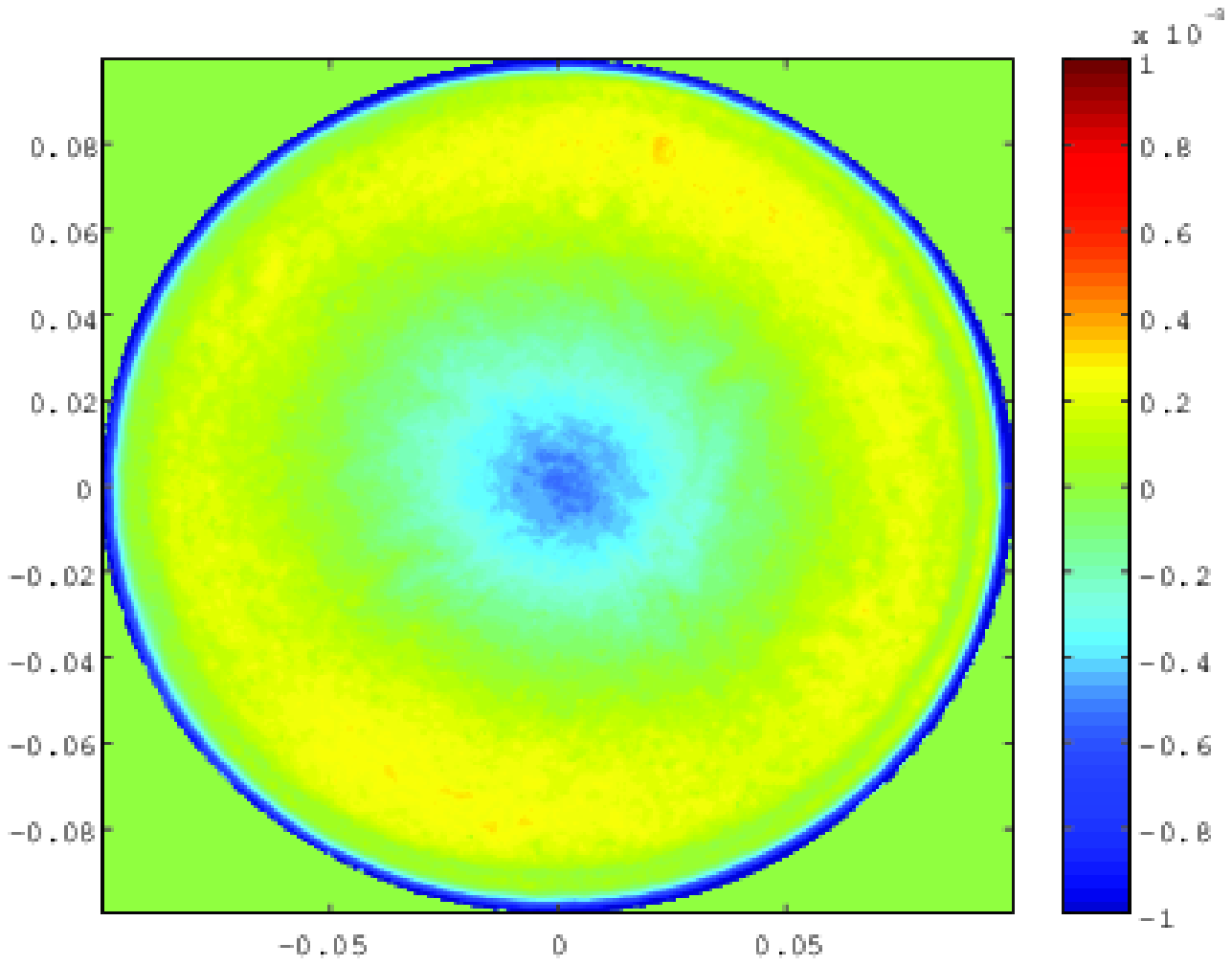}
		\includegraphics[width=4.25cm]{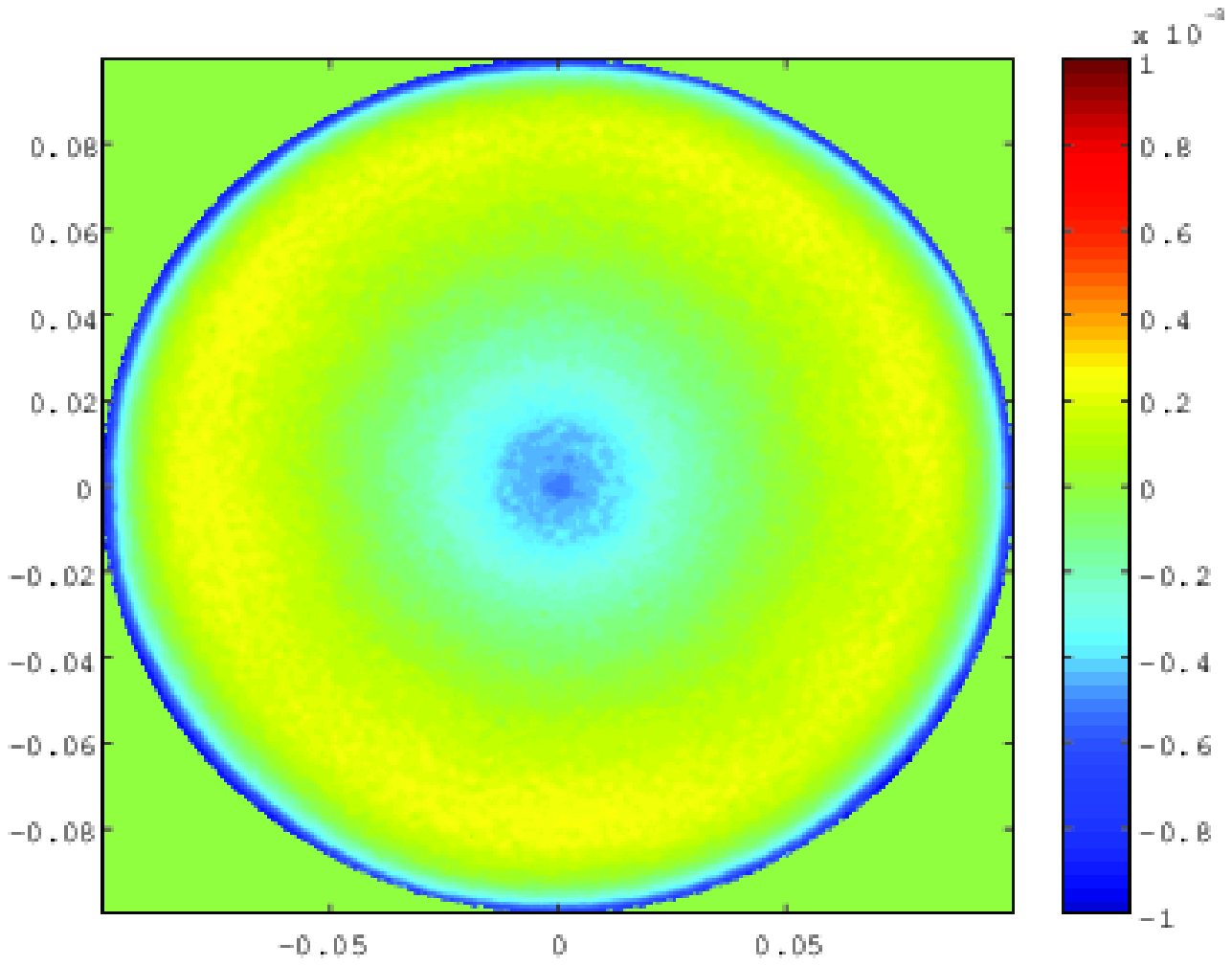}
		\includegraphics[width=4.25cm]{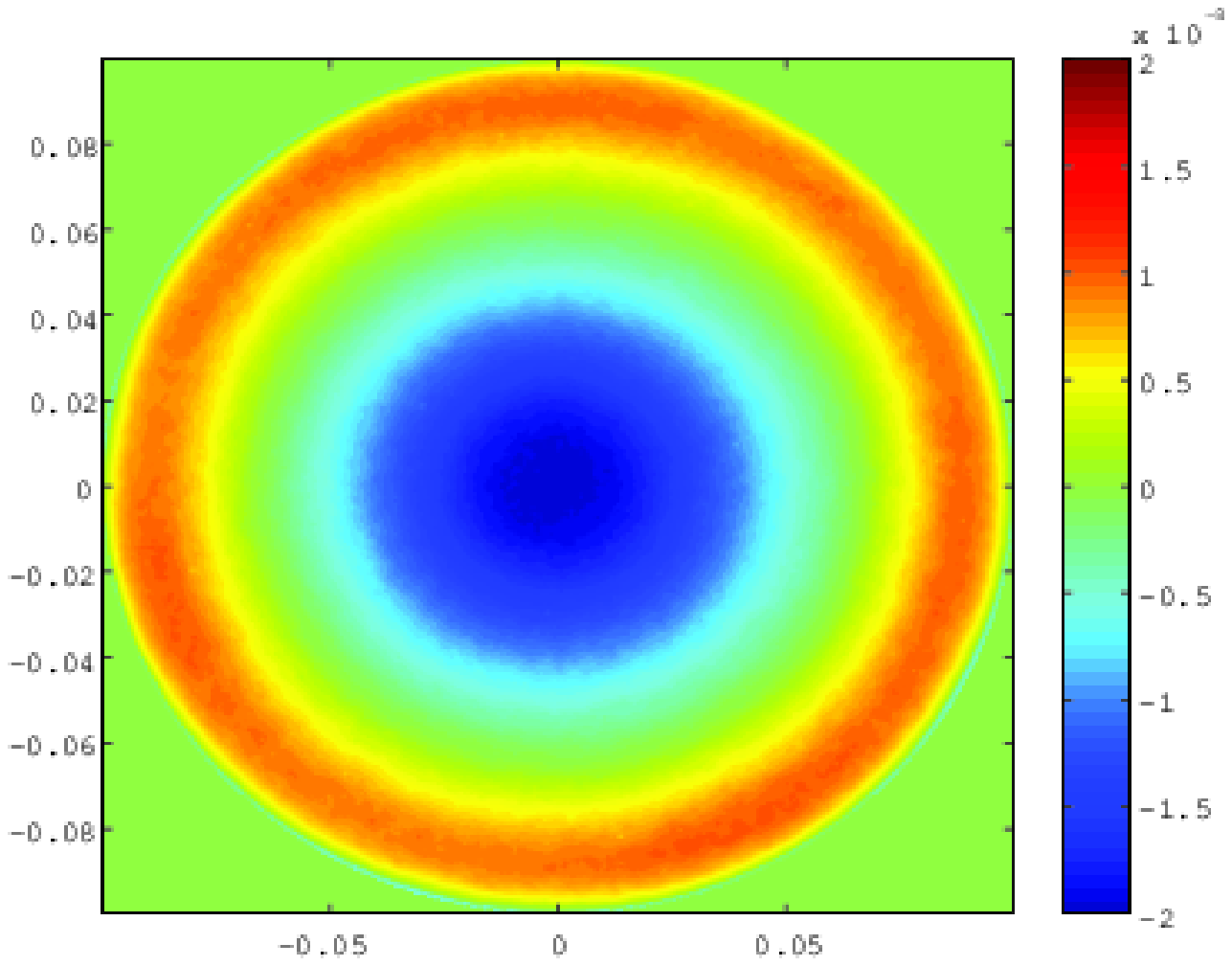}
		\includegraphics[width=4.25cm]{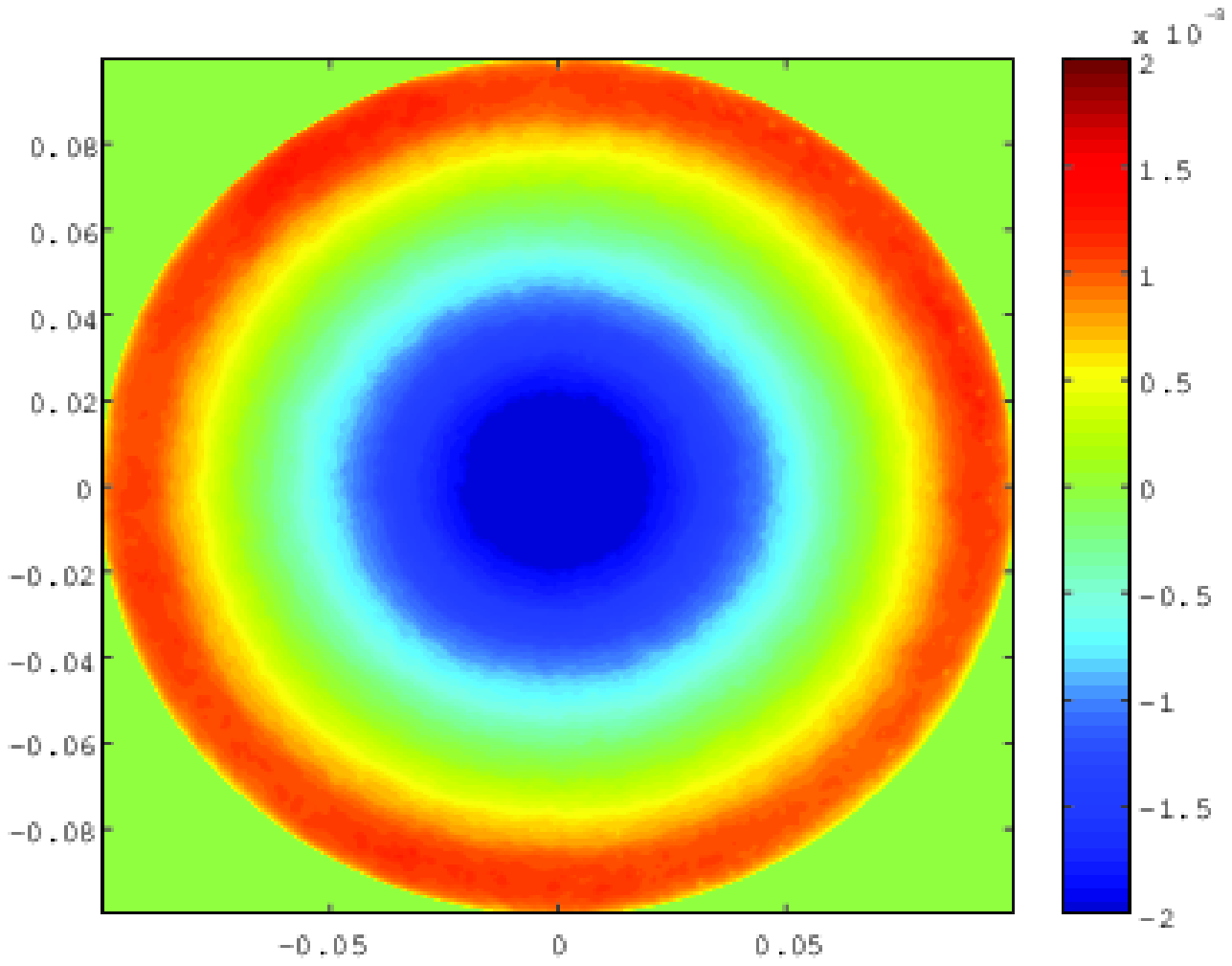}
	\caption{Phase maps of the high-reflective surface of the ITMX (top left), ITMY (top right), ETMX (bottom left) and ETMY (bottom right) after removing the power term, seen from the high-recflective coating side. The plotted area is 20\,cm in diameter around the center and the color bar ranges in $\pm1\,\mu\mathrm{m}$ for the input mirrors and in $\pm2\,\mu\mathrm{m}$ for the end mirrors.}
	\label{fig:HR}
	\end{center}
\end{figure}

In order to investigate the influence of the large TWE of the input mirrors, we need an interferometer simulation with the actual mirror maps. In this study, we use a simulation engine {\it FINESSE}, developed by Freise~\cite{FINESSE}. In the simulation, a light field is decomposed into the orthonormal set of Hermite-Gaussian modes. The amplitude of the beam is given by the mode order $(m,n)$, referring to the $x$- and $y$-directions. In the simulation, the fundamental mode, $(m,n)=(0,0)$, is defined by the fundamental mode of the input beam or by the mode of the cavity that the beam encounters last, and the beam is extended by the Hermite-Gaussian basis set. The highest order of Hermite-Gaussian polynomial is set by a user; in our simulation it is set to $N=12$. The beam parameter is transformed according to the respective ABCD matrix~\cite{Siegman} at each optical component. If a mirror map is implemented in an optical element, the phase and amplitude of the field are calculated in each segment defined in the map. While several kinds of mirror maps are provided in the FINESSE simulation, we use the reflection map and the transmission (TWE) map that give the phase shift at the reflection and the transmission, respectively. The measured TWE maps of the input mirrors are already shown in Fig.~\ref{fig:TWE}. The data points are $501\times501$. The measured reflection maps of the high reflective surface of the input mirrors and end mirrors (ETMX and ETMY) are shown in Fig.~\ref{fig:HR}. The data points are $501\times501$. Typically a mirror map has been pre-processed to remove average effects like a tilt of the entire surface or a curvature error so that the main players are the surface roughness and the non-uniformity of the refractive index.

The procedure of our simulation is as follows. First we build up an interferometer with radii of curvature properly given by analytical calculations. At this moment no mirror maps are implemented. We then define the cavities so that the mode basis sets are specified. The beam parameters of the input beam are given accordingly and they will not be changed afterwards. Then we replace the radii of curvature by the measured values and implement mirror maps. We also include an aperture on each test mass; 20\,cm in diameter. A mirror map causes a slight change of the average position of the reflection surface for the fundamental mode beam, and this change is modified manually. There are 7 steps to go. (i) To tune the position of the end mirror in each arm, we probe the transmitted beam to find the maximum with regard to the microscopic cavity length, or in other words the phase of the end mirror. For this simulation, the recycling cavities should be turned off; it is not only to set the reflectivities of the recycling mirrors to be zero but also to stop defining the recycling cavities. (ii) Once the end mirror positions are tuned, the next item to be tuned is the beam splitter position. This is necessary even though we do not implement the beam splitter map in this study. We probe the amplitude of the fundamental mode at the anti-symmetric port and find the minimum with regard to the phase of the beam splitter. Note that the output mode-cleaner is off at this point. (iii) Putting back the reflectivities of the recycling mirrors, we probe the transmitted beam of the end mirror of the x-arm cavity to find the maximum with regard to the power recycling mirror phase. (iv) In order to tune the phase of the signal recycling mirror, we need to generate a signal field. In our simulation we generate a differential signal at 10\,kHz. We probe the amplitude of the signal field at the anti-symmetric port without the output mode-cleaner and find the maximum with regard to the signal recycling mirror phase. In this study, we assume a non-detuned signal recycling cavity, which is the configuration of KAGRA in the first observation run. If one needs a simulation with the detuned signal recycling cavity, the detuned angle should be added to the phase of the signal recycling mirror after this tuning process. (v) Though not depicted in Fig.~\ref{fig:IFO}, there are two curved mirrors before the output mode-cleaner to mode-match the beam from the interferometer to the output mode-cleaner. We probe the amplitude of the 10\,kHz signal field at the anti-symmetric port with the output mode-cleaner and find maximum with regard to the distance of the two mode-matching telescope mirrors. In fact, this step is skipped for this study as we do not plan to move the curved mirrors for the mode matching at the first observation run in reality. (vi) Now we are to determine the amount of the offset for the reference beam to extract the gravitational-wave signal. The reference light consists of the component due to the imbalance of the reflectivities of the arm cavities and the component due to an intentional offset to the arm cavity lengths. While we plan to optimize the balance of these two components in the final stage of the KAGRA operation~\cite{KAGRA}, the offset component will be set much larger than the imbalance component in the first observation run. The phase of each arm cavity being finely tuned after process-(i), the reflectivity imbalance component, which is sometimes called the contrast defect, is the only fundamental-mode carrier light at the anti-symmetric port. If we like to tune the balance between the two components, we measure the power of the contrast defect and then determine the amount of offset. If we like the reference light to be mostly the offset component, the amount of the offset is determined to make the total light power at the anti-symmetric port be $\sim40$\,mW. In this study, we have chosen the latter setup. (vii) The last step is to align the beam into the output mode-cleaner. This step would not be necessary if the mirror map did not have an asymmetric structure to create a virtual tilt on the reflected/transmitted beam. Indeed the beam was already centered to the output mode-cleaner in our previous studies with artificial mirror maps~{\cite{Kumeta}\cite{Yano}}. Instead of tilting the four-mirror bowtie cavity simultaneously in the FINESSE simulation, we tilt the pair of steering mirrors before the output mode-cleaner to maximize the transmission of the signal field. Table~\ref{tab:IFO} shows a list of major interferometer parameters.

\begin{table}[htbp]
\begin{center}
\begin{tabular}{c|c}
item&value\\ \hline
Laser power&55.375\,W\\
Control sideband frequencies&16.88\,MHz / 45.01\,MHz\\
BS-ITMX distance&26.6649\,m\\
BS-ITMY distance&23.3351\,m\\
Recycling cavity length&66.59\,m\\
ITMX/Y RoC&1904.6\,m\,/\,1904.4\,m\\
ETMX/Y RoC&1908.24\,m\,/\,1905.55\,m\\
ITMX/Y reflectivities&99.602\,\%\,/\,99.598\,\%\\
ETMX/Y reflectivities&99.9941\,\%\,/\,99.9949\,\%\\
PRM/SRM reflectivity&90\,\%\,/\,70\,\%\\
OMMT mirrors RoC&13.16\,m / 44.39\,m\\
OMC Gouy phase&55.4\,deg\\
OMC finesse&780
\end{tabular}
\caption{Some interferometer parameters used in the simulation. The radii of curvature (RoC) of the test masses are the ones measured~\cite{HiroseInPrep}. The reflectivities of the test masses are given with an assumption of 10\,\% imbalance in the loss and 0.5\,\% imbalance in the finesse. Here, PRM, SRM, OMMT, and OMC stand for the power recycling mirror, signal recycling mirror, the output mode-matching telescope, and the output mode-cleaner, respectively.}
\label{tab:IFO}
\end{center}
\end{table}

\section{Results}\label{sec:5}

\subsection{Parameters after the tuning}\label{sec:5-0}

As mentioned in Sec.~\ref{sec:4}, the phases of some mirrors have to tuned after implementing mirror maps. We performed two simulations: one is only with the HR maps of the test masses and the other is with the HR maps and the TWE maps of the input mirrors. Table~\ref{tab:tuned} shows the mirror phases after the tuning process. The output mode-matching telescope length was not changed after the tuning with the mirror maps. 
The output mode-cleaner tilt and centering did not change much with the HR maps. It did not change much even with the TWE maps in the vertical direction ($y$-direction).

\begin{table}[htbp]
\begin{center}
\begin{tabular}{c|c|c|c}
&no map&HR only&HR+TWE\\ \hline
ETMX phase (deg)&90.0012&96.3419&96.3420\\
ETMY phase (deg)&-0.0012&6.9570&6.9569\\
BS phase (deg)&0&0.2098&8.5733\\
PRM phase (deg)&0&0.4604&8.9757\\
SRM phase (deg)&0&-0.9631&2.0147\\
OMMT length (m)&2.6&2.6&2.6\\
OMC $x$-tilt (rad)&0&$-12\mu$&$-96\mu$\\
OMC $x$-centering (m)&0&$7\mu$&$13\mu$\\
OMC $y$-tilt (rad)&0&$0\mu$&$-4\mu$\\
OMC $y$-centering (m)&0&$0\mu$&$0\mu$\\
\end{tabular}
\caption{Mirror phases after the tuning with the HR maps only and with the HR+TWE maps. See the caption of Table \ref{tab:IFO} for the abbreviations.}
\label{tab:tuned}
\end{center}
\end{table}

\subsection{Higher order modes at the anti-symmetric port}\label{sec:5-1}

\begin{figure}[htbp]
	\begin{center}
		\includegraphics[width=4cm]{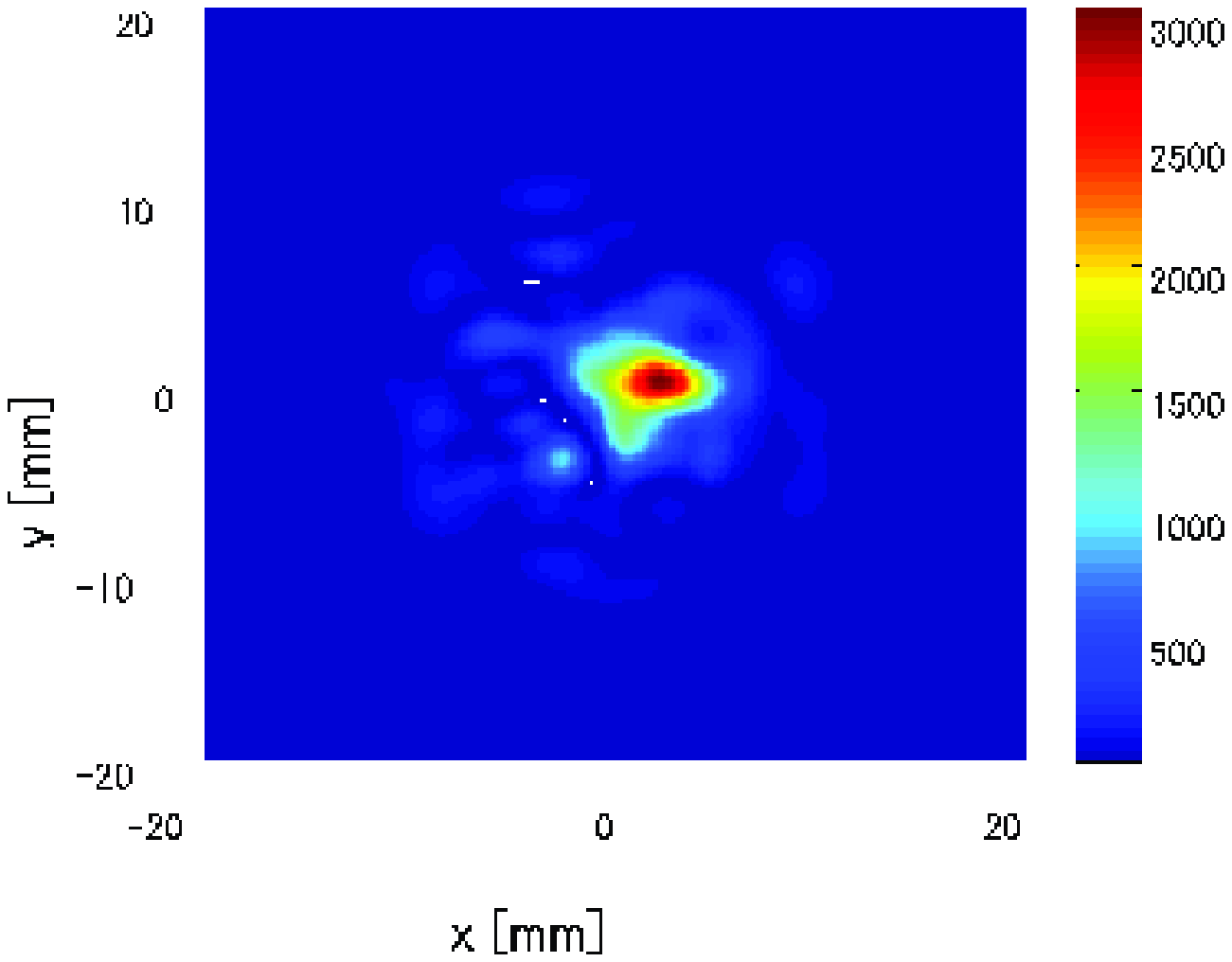}
		\includegraphics[width=4cm]{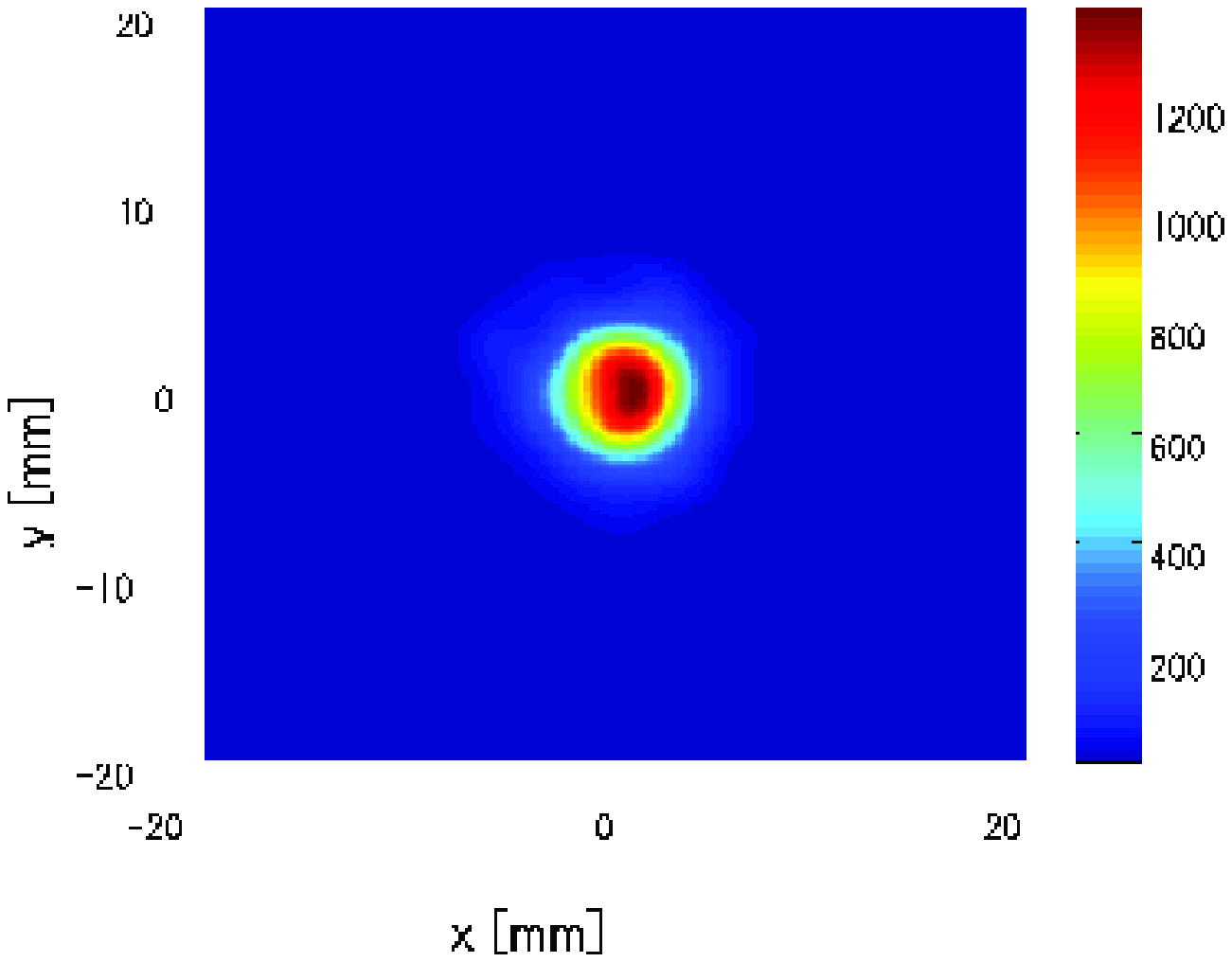}
	\caption{The images of the beam right after the signal recycling mirror with and without the input mirror substrate transmitted wavefront error ({\it left} and {\it right}, respectively).}
	\label{fig:beam}
	\end{center}
\end{figure}

Figure~\ref{fig:beam} shows the beam images right after the signal recycling mirror. The color bar shows the intensity in the unit of $\mathrm{W}/\mathrm{m}^2$. The control sideband fields are removed. The fundamental mode carrier beam is not removed. One can see that the TWE maps have made the beam dirty, or in other words more spatial higher order modes mixed in the fundamental mode.

The output mode-cleaner can filter out the higher order modes to some extent. Table~\ref{tab:modes} shows the power of each mode before and after the output mode-cleaner in the case with and without the input mirror TWE maps. Note that we do not assume any optical losses or any surface distortion in the output mode-cleaner. A contribution of the optical loss in the actual output mode-cleaner transmission is roughly estimated to be $\sim2\,\%$. One can see that the higher order modes increased with the TWE maps by one or two orders of magnitude. The total power of the higher order modes after the output mode-cleaner is 0.11\,$\mu$W without the TWE and 1.82\,$\mu$W with the TWE; it is 16 times more with the TWE but is as low as 40\,ppm of the reference light thanks to the high finesse output mode-cleaner.

\begin{table}[htbp]
\begin{center}
\begin{tabular}{|c|c|c|c|c|}
\hline
&\multicolumn{2}{|c|}{HR only (W)}&\multicolumn{2}{|c|}{HR+TWE (W)}\\ \hline
$n$&before OMC&after OMC&before OMC&after OMC\\ \hline
0&4.3e-2&4.0e-2&4.3e-2&4.7e-2\\
1&1.7e-3&6.2e-9&1.8e-2&2.3e-8\\
2&1.1e-3&2.4e-9&1.1e-2&3.6e-8\\
3&4.0e-4&2.7e-8&4.7e-3&3.0e-7\\
4&4.8e-4&3.6e-9&6.0e-3&7.9e-8\\
5&2.2e-5&9.6e-11&3.8e-3&2.1e-8\\
6&5.7e-5&1.4e-9&5.5e-3&1.4e-7\\
7&1.6e-4&2.9e-9&1.4e-3&2.8e-8\\
8&1.0e-4&8.3e-10&1.0e-2&4.1e-8\\
9&1.9e-4&1.2e-9&1.3e-2&1.1e-7\\
10&8.4e-4&3.5e-8&2.3e-2&6.3e-7\\
11&1.1e-3&2.7e-8&3.1e-2&3.2e-7\\
12&8.5e-4&5.2e-9&2.0e-2&7.7e-8\\ \hline
\end{tabular}
\caption{Power of $n$-th order modes ($n=0,1,2\cdots,12$) before and after the output mode-cleaner with and without the TWE maps.}
\label{tab:modes}
\end{center}
\end{table}

Here the fundamental mode beam is not exactly 40\,mW in some cases. This is because the step (vii) was taken after tuning the offset. The output mode-cleaner cavity mode was redefined and the fundamental mode increased. 

\subsection{Shot-noise-limited sensitivity}\label{sec:5-2}

An increase of shot noise due to the higher order modes will be about $0.7\,\%$ or less with the $\sim40$\,mW reference light. If we reduce the reference light power to $\sim10$\,mW, the contribution will be $1.3\,\%$. 

A larger concern would be a possible decrease of the signal field after the output mode-cleaner. If the wavefront of the signal field is distorted by the transmission error of the input mirrors in a non-spherical way, a fraction of the signal field beam will have to be filtered by the output mode-cleaner. (In fact, as we skipped the step (iv), even a spherical distortion cannot be compensated by tuning the output mode-matching telescope in this study.) We performed a simulation to calculate the signal field amplitude of each mode and the contribution of the higher order modes in the total signal field at the anti-symmetric port was $0.08\,\%$ with the HR maps only and $4.5\,\%$ with the HR+TWE maps.

Another concern would be a possible decrease of the power recycling gain. We performed a simulation to compare the power recycling gain of the fundamental mode and it was 9.4 with the HR maps only and 8.5 with the HR+TWE maps. Note that we do not assume any optical losses in the power recycling mirror or in the beam splitter.

\begin{figure}[htbp]
	\begin{center}
		\includegraphics[width=7cm]{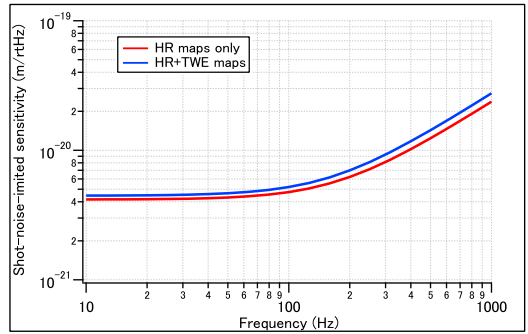}
	\caption{Shot-noise-limited sensitivities with the HR maps only and with the HR+TWE maps.}
	\label{fig:shot}
	\end{center}
\end{figure}

Figure~\ref{fig:shot} shows the shot-noise-limited sensitivities with the HR maps only and with the HR+TWE maps. The shot noise level is 7.1\,\% worse at low frequencies and 16\,\% worse at high frequencies with the TWE maps.

\subsection{Laser noise coupling}\label{sec:5-3}

\begin{figure}[htbp]
	\begin{center}
		\includegraphics[width=7cm]{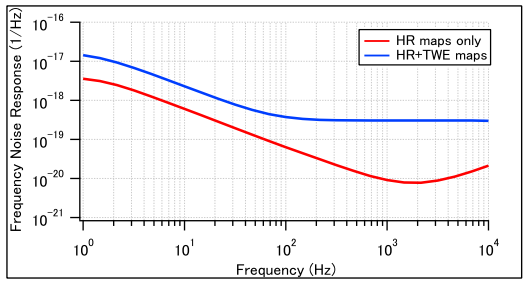}
	\caption{Laser noise couplings with and without the TWE maps.}
	\label{fig:lasernoise}
	\end{center}
\end{figure}

All the second-generation gravitational-wave telescopes including KAGRA employ the DC readout scheme, which is to send a reference beam at the carrier frequency to the anti-symmetric port and take a beat note with the signal field. This is an updated readout scheme from the conventional RF readout scheme, which is to modulate the input field by a radio frequency and take a beat note of this sideband field and the signal field at the anti-symmetric port. One advantage of the DC readout scheme is that the reference beam, mainly generated with a differential offset to the arm length, experiences both a low-pass filter of the arm cavity and a low-pass filter of the power-recycling cavity to suppress a coupling of laser noise~\cite{SomiyaPRD}. If the reference beam is not generated in the arm cavities but in the central interferometer, the second low-pass filter will be eliminated and the laser noise contribution will increase. We perform a simulation to investigate the laser noise coupling with the substrate's non-uniformity. Figure~\ref{fig:lasernoise} shows the result. One can see that the laser frequency noise coupling increases with the TWE maps more than one order of magnitude in the observation frequency band.

\subsection{Wave-front-sensing signal}\label{sec:5-4}

A possible distortion of a control signal is another type of concern with this kind of structure in the transmissive optics. We calculated the wave-front-sensing signal of the common-mode tilt of the arm cavities obtained at the symmetric port. The wave-front-sensing is a well-established method to probe a small tilt of optics in a cavity~\cite{WFS}. A TEM10 or TEM01 mode of the probe beam in a cavity couples with a reference beam, which is usually a radio-frequency sideband field that does not enter the cavity, to be detected at a pair of QPDs. The tilts of the cavity optics can be decomposed into a symmetric (hard) mode and anti-symmetric (soft) mode. Each mode of the two arm cavities makes common and differential modes. The Gouy phases of the beam at the two QPDs are properly chosen to minimize the cross coupling of the signals. 

\begin{figure}[htbp]
	\begin{center}
		\includegraphics[width=8.5cm]{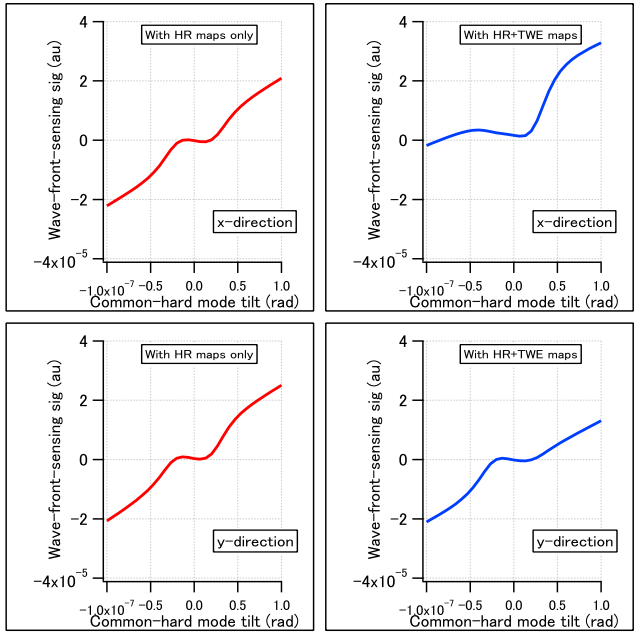}
	\caption{{\it Top}: Wave-front-sensing signals in $x$-direction with the HR maps only (left) and with the HR+TWE maps (right). {\it Bottom}: Wave-front-sensing signals in $y$-direction with the HR maps only (left) and with the HR+TWE maps (right).}
	\label{fig:WFS}
	\end{center}
\end{figure}

The top and bottom panels of Fig.~\ref{fig:WFS} shows the wave-front-sensing signals with regard to the common-hard mode tilt, as an example, in the $x$-direction and in the $y$-direction, respectively. Compared with the case without the TWE maps, the wave-front-sensing signal in the $x$-direction is deformed and the zero-crossing point is offset with the TWE maps.

\subsection{Alignment control of the output mode-cleaner}\label{sec:5-5}

The alignment control signal of the output mode-cleaner is obtained from the PD at the transmission of the cavity and from the QPDs at the pickoff before the cavity. The former is for a fine alignment and the latter is for a course alignment. All the photo detectors are mounted on a semi-monolithic breadboard for the output mode-cleaner so the QPDs can probe a relative tilt and miscentering of the beam with regard to the output mode-cleaner cavity. The QPDs are about 46\,cm distant where the Rayleigh range is about 56\,cm.

\begin{table}
\begin{minipage}{4cm}
\begin{tabular}{|p{10mm}|p{13mm}|p{13mm}|}
\multicolumn{3}{c}{HR only}\\ \hline
&$x$-misc&$x$-tilt\\ \hline
QPD1&-9.7\,{\tiny W/m}&-1.0\,{\tiny W/rad}\\ \hline
QPD2&8.0\,{\tiny W/m}&4.5\,{\tiny W/rad}\\ \hline \hline
&$y$-misc&$y$-tilt\\ \hline
QPD1&-9.8\,{\tiny W/m}&1.0\,{\tiny W/rad}\\ \hline
QPD2&-8.1\,{\tiny W/m}&4.6\,{\tiny W/rad}\\ \hline
\end{tabular}
\end{minipage}
\begin{minipage}{4cm}
\begin{tabular}{|p{10mm}|p{13mm}|p{13mm}|}
\multicolumn{3}{c}{HR+TWE}\\ \hline
&$x$-misc&$x$-tilt\\ \hline
QPD1&-8.3\,{\tiny W/m}&-1.9\,{\tiny W/rad}\\ \hline
QPD2&6.0\,{\tiny W/m}&4.3\,{\tiny W/rad}\\ \hline \hline
&$y$-misc&$y$-tilt\\ \hline
QPD1&-7.6\,{\tiny W/m}&1.3\,{\tiny W/rad}\\ \hline
QPD2&-7.3\,{\tiny W/m}&4.4\,{\tiny W/rad}\\ \hline
\end{tabular}
\end{minipage}
\caption{Sensing matrices of the alignment control of the output mode-cleaner with and without the TWE maps.}
\label{tab:qpd}
\end{table}

Table~\ref{tab:qpd} shows the sensing matrices of the alignment control of the output mode-cleaner. Two QPDs probe the tilt and miscentering. The control signals can be obtained separately as far as the responses to the two degrees of freedom are somewhat independent. According to Ref.~\cite{SomiyaAO}, the sensing noise level of a 2-dof system with a sensing matrix $((a,\,b),\,(c,\,d))$ is given by
\begin{eqnarray}
\langle x_1 \rangle&=&\frac{1}{1-\eta}\left(\frac{\langle n_1\rangle}{a}+\eta\frac{\langle n_2\rangle}{c}\right)\,,\nonumber\\
\langle x_2 \rangle&=&\frac{1}{1-\eta}\left(\frac{\langle n_2\rangle}{d}+\eta\frac{\langle n_1\rangle}{b}\right)\,,\nonumber
\end{eqnarray} 
with the degeneracy factor $\eta\equiv bc/ad$. In our case, $\langle x_1 \rangle$ and $\langle x_2 \rangle$ are miscentering sensing noise and tilt sensing noise, respectively; the bracket means the ensemble average. Here $\langle n_j\rangle$ is noise on a photo detector, which is proportional to the square root of the light power on a photo detector, hence almost the same with or without the TWE maps. For an effective tilt/miscentering sensing with two QPDs, the distance of the QPDs is set far enough compared with the Rayleigh range and the degeneracy factor $\eta$ is far from unity without the TWE maps: 0.18 for the miscentering and 0.18 for the tilt sensing. With the TWE maps included, however, the degeneracy factor becomes 0.32 for the miscentering and 0.28 for the tilt sensing.

\subsection{Tilt coupling at the output mode-cleaner}\label{sec:5-6}

Through a design study of KAGRA's output mode-cleaner, we have found that an asymmetric structure in a transmissive object results in a severer requirement on the seismic isolation of the output mode-cleaner. As explained in Sec.~\ref{sec:5-3}, we need a reference light to extract a gravitational-wave signal. A tilt or miscentering of the output mode-cleaner generates an amplitude modulation on the reference light, converting the fundamental mode into first-order Hermite-Gaussian modes. This effect could be, however, quadratic to the tilt or miscentering, if the cavity is well aligned to the reference light. Incidentally, the tilt or miscentering also causes a reduction of the gravitational wave signal transmitting through the output mode-cleaner. This effect could be also quadratic to the tilt or miscentering. In fact, the optimal alignment to maximize the signal transmission is different from that to minimize the conversion of the reference light if one of the transmissive objects has an asymmetric structure.

\begin{figure}[htbp]
	\begin{center}
		\includegraphics[width=8.0cm]{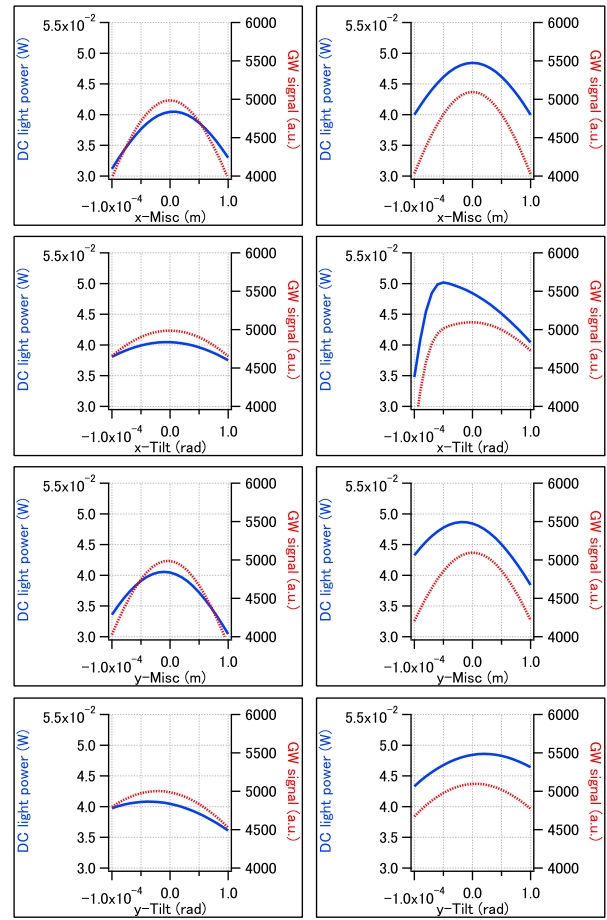}
	\caption{The power of reference DC light power (solid curve) and the gravitational-wave (GW) signal field (dashed curve) transmitting through the output mode-cleaner with regard to the tilt and miscentering of the output mode-cleaner in $x$ and $y$ directions. {\it Left}: with HR maps only. {\it Right}: with HR + TWE maps.}
	\label{fig:OMCcoupling}
	\end{center}
\end{figure}

Figure~\ref{fig:OMCcoupling} shows how much the reference light power changes with the tilt and miscentering of the output mode-cleaner. With the HR maps only, the reference light power mostly peaks at the operation point where the signal transmission is maximized. With the TWE maps included, however, the reference light peaks at a different point for the tilt in the $x$-direction. The response to the tilt in the $x$-direction at the operation point is calculated to be $52\,\mathrm{W/rad}$. It is 16 times larger than that in the $x$-direction with the HR maps only. The response to the tilt is not as good in the $y$-direction even with the HR maps only; $20\,\mathrm{W/rad}$, but the response in the $x$-direction with the TWE maps is even worse. A large response imposes a severe requirement on the seismic isolation system of the output mode-cleaner. 

\section{Summary and discussions}\label{sec:6}

In this paper, we performed a number of simulations with measured mirror maps to investigate an influence of the non-uniformity in the sapphire substrate used as an input mirror of KAGRA. First we calculated the amount of the spatial higher-order modes at the anti-symmetric port and found that the total amount of the higher-order modes power is 16 times larger with the transmission maps. It turned out that 4.5\,\% of the signal field scatters away to the higher order modes with the TWE maps while it is only 0.08\,\% with the HR maps only. We also found that the power recycling gain was reduced from 9.4 to 8.5. As a result the shot noise level increased for 7.1\,\% at low frequencies and 16\,\% at high frequencies. The laser frequency noise level increased by more than one order of magnitude in the observation band. The wave-front-sensing signal of the common hard mode is deformed especially in the x-direction with the TWE maps. The degeneracy of the output mode-cleaner alignment sensing matrix is increased by a factor of $1.5\sim2$. At last, we found that the output mode-cleaner tilt noise coupling in the $x$-direction increases with the TWE maps. It is 2.6 times larger than that in the $y$-direction with the HR maps only, which is the largest before introducing the TWE maps.

Among all, the severest issue would be the laser noise coupling. According to Ref.~\cite{SomiyaPRD}, there is a few orders of magnitude of the safety margin between the frequency-noise-limited sensitivity and the quantum-noise-limited sensitivity of the broadband configuration. We assume, however, a 0.5\,\% imbalance in the arm cavity finesse and losses. A recent measurement of the input mirror reflectivity has revealed a larger ($\sim10\,\%$) imbalance due to a non-simultaneous coating production. This additional factor of 20 would bring frequency noise quite close to limit the sensitivity of KAGRA. A further investigation with the latest parameters is required.

It has been reported from the LIGO group that a point defect in one of the silica substrates locally increases the temperature with absorbing the heat from the laser and causes a wavefront distortion~\cite{LIGOdefect}. The phenomenon is analogous to the non-uniformity of the sapphire substrate in KAGRA. They have decided to displace the beam from the center of the mirror so that the heat absorption at the point defect decreases. KAGRA could try the same but the effect would be less as our structure in the TWE map exists even if the laser power is reduced. A further investigation using the modal model simulation is required to see how much improvement can be expected in KAGRA with the intentional miscentering of the beam.

This paper demonstrates a usage of the modal model simulation to investigate an influence of some defects in a gravitational wave telescope. This kind of simulation is a powerful tool to design the instrument before the actual start of an observation and also to characterize observed phenomena at the commissioning of the instruments during the observation run.

\begin{acknowledgements}
This work was supported by MEXT, JSPS Leading-edge Research Infrastructure Program, JSPS Grant-in-Aid for Specially Promoted Research 26000005, JSPS Grant-in-Aid for Scientific Research on Innovative Areas 2905: JP17H06358, JP17H06361 and JP17H06364, JSPS Core-to-Core Program A. Advanced Research Networks, the joint research program of the Institute for Cosmic Ray Research, University of Tokyo, and the Mitsubishi Foundation. This research activity has also been supported by the European Commission within the Seventh Framework Programme (FP7)-Project ELiTES (GA 295153). 
The authors would like to express special thanks Hiro Yamamoto at Caltech for valueable discussions.
\end{acknowledgements}

\bibliographystyle{junsrt}
\pagestyle{headings}

\end{document}